\documentclass[a4paper,11pt]{article}
\pdfoutput=1 

\usepackage{jcappub} 

\usepackage[T1]{fontenc} 
\usepackage{booktabs}
\usepackage{mathtools}
\usepackage{multirow}
\usepackage{float}
\usepackage{tabularx} 
\usepackage{array}  
\usepackage[left]{lineno}

\title{Search for solar atmospheric neutrinos with the ANTARES neutrino telescope}

\collaboration{\includegraphics[width=0.13\textwidth]{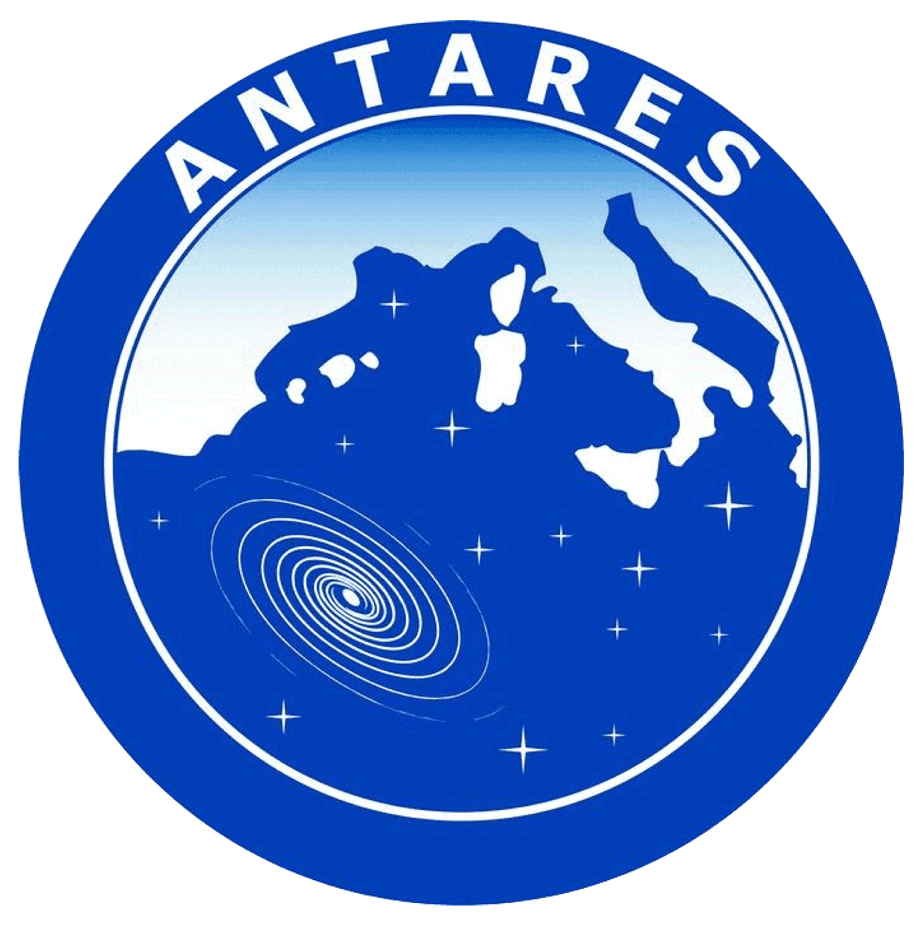}\vspace*{0.2cm}
\\The ANTARES collaboration\\}

\emailAdd{daniellc@ugr.es}
\author[1,2]{A.~Albert}
\author[3]{S.~Alves}
\author[4]{M.~Andr\'e}
\author[5]{M.~Anghinolfi}
\author[6]{G.~Anton}
\author[7]{M.~Ardid}
\author[7]{S.~Ardid}
\author[8]{J.-J.~Aubert}
\author[9]{J.~Aublin}
\author[9]{B.~Baret}
% ...

\affiliation[1]{\scriptsize{Universit\'e de Strasbourg, CNRS,  IPHC UMR 7178, F-67000 Strasbourg, France}}
\affiliation[2]{\scriptsize Universit\'e de Haute Alsace, F-68100 Mulhouse, France}
\affiliation[3]{\scriptsize{IFIC - Instituto de F\'isica Corpuscular (CSIC - Universitat de Val\`encia) c/ Catedr\'atico Jos\'e Beltr\'an, 2 E-46980 Paterna, Valencia, Spain}}
\affiliation[4]{\scriptsize{Technical University of Catalonia, Laboratory of Applied Bioacoustics, Rambla Exposici\'o, 08800 Vilanova i la Geltr\'u, Barcelona, Spain}}
\affiliation[5]{\scriptsize{INFN - Sezione di Genova, Via Dodecaneso 33, 16146 Genova, Italy}}
\affiliation[6]{\scriptsize{Friedrich-Alexander-Universit\"at Erlangen-N\"urnberg, Erlangen Centre for Astroparticle Physics, Erwin-Rommel-Str. 1, 91058 Erlangen, Germany}}
\affiliation[7]{\scriptsize{Institut d'Investigaci\'o per a la Gesti\'o Integrada de les Zones Costaneres (IGIC) - Universitat Polit\`ecnica de Val\`encia. C/  Paranimf 1, 46730 Gandia, Spain}}
\affiliation[8]{\scriptsize{Aix Marseille Univ, CNRS/IN2P3, CPPM, Marseille, France}}
\affiliation[9]{\scriptsize{Universit\'e de Paris, CNRS, Astroparticule et Cosmologie, F-75013 Paris, France}}
% ...

\abstract{Solar Atmospheric Neutrinos (SA$\nu$s) are produced by the interaction of cosmic rays with the solar medium. The detection of SA$\nu$s would provide useful information on the composition of primary cosmic rays as well as the solar density. These neutrinos represent an irreducible source of background for indirect searches for dark matter towards the Sun and the measurement of their flux would allow for a better assessment of the uncertainties related to these searches. In this paper we report on the analysis performed, based on an unbinned likelihood maximisation, to search for SA$\nu$s with the ANTARES neutrino telescope. After analysing the data collected over 11 years, no evidence for a solar atmospheric neutrino signal has been found. An upper limit at 90\% confidence level on the flux of solar atmospheric neutrinos has been obtained, equal to 7$\times$$10^{-11}$ [ TeV$^{-1}$cm$^{-2}$s$^{-1}$] at E$_\nu =$ 1 TeV for the reference cosmic ray model assumed.}


\begin{document}
\maketitle
\flushbottom

\section{Introduction}
\label{sec:Introduction}
Apart from the electromagnetic radiation, the Sun is also a source of neutrinos.\footnote{Hereafter, the word \emph{neutrino} refers to both, $\nu$ and $\bar{\nu}$ unless otherwise specified.} 
An intense flux of neutrinos is generated by nuclear fusion reactions in the Sun core at MeV energies \cite{Antonelli2013, Bahcall1989, Haxton2013}. At higher energies (GeV to TeV), an additional flux of neutrinos is expected from the Sun direction, coming from the decay products of the Cosmic Rays (CRs) interacting in the Sun. The flux of these neutrinos, so-called Solar Atmospheric Neutrinos (SA$\nu$s), would then reach the Earth \cite{Moskalenko_1993,masip_high_2018} after being modulated by oscillation phenomena. SA$\nu$s represent an unavoidable source of background for Dark Matter (DM) indirect searches \cite{adrianmartinez_limits_2016, AdrianMartinez2016, Ng2017, argueelles_solar_2017, navas_dark_2019}, even though the sensitivity of current experiments has not yet been able to determine the intensity of their flux. The detection of SA$\nu$s would allow for the characterisation of this potential background. In addition, it also can shed light on understanding of the primary CR composition, the solar density or even the parameters of neutrino oscillation \cite{edsjoe_neutrinos_2017}. In addition, a better understanding of the gamma-ray flux arising from the CR interaction in the solar atmosphere would be important to have a robust prediction of the SA$\nu$ flux \cite{Tang_2018, Linden2020}.

Even though the production mechanism of SA$\nu$s is similar to the one of the terrestrial atmospheric neutrinos, the flux of SA$\nu$s is expected to be slightly harder. Indeed, since  the solar atmosphere is less dense than the Earth's atmosphere, the secondary particles pro\-du\-ced by the CR interaction are more likely to decay than to interact in the solar medium. Considering that at sufficiently large depth in the solar medium almost every secondary cascade would have decayed, the overall neutrino production occurs on the solar surface. The flux ratio at production site is approximately $\{ \nu_e \}: \{ \nu_\mu \}: \{ \nu_\tau \} = 1:2:0$. However, the final neutrino flux at Earth, after oscillations, has a flavour ratio of $1:1:1$ \cite{edsjoe_neutrinos_2017}. Since the solar density and composition play a crucial role in the final neutrino flux (changing the production flux up to a $\pm30\%$ \cite{edsjoe_neutrinos_2017}), several models have been proposed in the literature over the years \cite{serenelli_new_2009, Hettlage_2000,Grevesse_1998, Seckel1991}.

In this paper, a search for SA$\nu$s using 11 years of ANTARES data (2008--2018) is presented.  The outline of the paper is the following: in section \ref{sec:ANTARES}, the ANTARES detector, the event topologies and the main background sources present in a neutrino telescope are des\-cribed. The ANTARES simulation and reconstruction chain, as well as the event selection and the different models tested in the analysis, are presented in section \ref{sec:data_set}. Section \ref{sec:Analysis} describes the strategy used to search for an excess of SA$\nu$s over the background. Finally, the results and conclusions of the analysis are presented and discussed in section \ref{sec:Results}. 

\section{The ANTARES neutrino telescope}
\label{sec:ANTARES}

The ANTARES detector is anchored at a depth of $\sim$ 2500~m on the Mediterranean seabed, 40~km offshore from Toulon \cite{ageron_antares_2011}. The ANTARES first detection line was deployed in 2006, and the apparatus reached its full configuration in 2008. Since then, the ANTARES detector has been taking data almost continuously.

ANTARES is made of 12 detection lines of 450 m height, spaced by about 60--75 m and distributed on an octogonal grid. Each line holds 25 storeys vertically spaced by about 14 m. Each storey hosts 3 optical modules (OMs), made of high-pressure resistant glass transparent to 400--500 nm photons, and a local control module containing the electronics \cite{aguilar_performance_2010, aguilar_data_2007}. The 12$^\textnormal{th}$ line has 20 storeys equipped with OMs, and 5 storeys with acoustic devices for neutrino detection \cite{aguilar_amadeusacoustic_2011} and additional instruments to measure environmental parameters \cite{Aguilar_2006}. 

The main component of the OM \cite{amram_antares_2002} is a 10" photomultiplier tube (PMT), oriented 45$^{\circ}$ downward in order to optimise the detection of light induced by relativistic upgoing charged particles, and to mitigate the effect of sedimentation and biofouling \cite{Amram_2003}. A $\mu$-metal cage is used in order to shield the electron trajectory inside the PMT from the effect of the Earth magnetic field.

Photons impinging in the sensitive area of a PMT generate a signal called \emph{``hit''} \cite{aguilar_performance_2010}, which carries information of the arrival time and collected charge. The information of all recorded hits in the detector is used to reconstruct the direction and energy of the event. Apart from the light induced by charged particles yielded by the neutrino interaction in the detector surroundings (the signal), other background light sources are present, the dominant ones being: atmospheric downgoing muons, bioluminescence activity and $^{40}$K decays \cite{Amram_2000}. The influence of the environmental background over the overall signal is reduced by applying dedicated trigger algorithms \cite{ageron_antares_2011}.

Depending on the interaction process, neutrino induced events fall into two main ca\-te\-go\-ries: \emph{track-like} and \emph{shower-like} events. Charged Current (CC) interactions of $\nu_\mu$ and CC interactions of $\nu_\tau$ with a subsequent muonic decay produce a long-range muon, the so-called \emph{track-like} event topology. All other neutrino interactions, both CC and Neutral Current (NC), produce a cascade of charged particles, and are classified as \emph{shower-like} events.

The ANTARES median angular resolution for muon \emph{tracks} ranges from $1^{\circ}$ for energies below 1 TeV, to better than $0.4^{\circ}$ for neutrino energies above 10 TeV \cite{ageron_antares_2011}. 

\section{Simulation, reconstruction and event selection
\label{sec:data_set}}

In this section the generation of the signal due to the SA$\nu$, the description of the background, the reconstruction of tracks, and the event selection are described.

\subsection{SA$\nu$ generation}

The signal in this analysis is represented by SA$\nu$s produced by CR interactions in the Sun. The expected neutrino flux depends on the used CR flux and interaction models, and on the spatial distribution of the incoming neutrinos on Earth. Effects on oscillations must also be considered.

The \emph{Hillas-Gaisser 3-generation model (H3a)} \cite{Gaisser_2012} and the \emph{Gaisser-Stanev-Tilav 4-generation model (GST4)} \cite{Gaisser_2013} are chosen as input CR models.
Concerning the solar density profiles, the \emph{Ser+Stein} \cite{serenelli_new_2009} and the \emph{Grevesse \& Sauval} (referred to as \emph{Ser+GS98} \cite{Grevesse_1998}) are used. 

All these models are available in the \emph{WimpSim 5.0} simulation framework \cite{edsjoe_wimpsim_2019, blennow_neutrinos_2008}, 
which includes tools in the \texttt{solar\_crnu} package to calculate the neutrinos arising from CR interactions in the solar atmosphere. The detail of hadronic interactions are simulated with the \texttt{MCEq}  code\footnote{See: A. Fedynitch et al, https://github.com/afedynitch/MCEq.}. In \emph{WimpSim}, the \texttt{MCEq} code has been run to create data files that give the production fluxes for different energies, impact parameters and depth in the solar atmosphere. 
For more details, including discussion to systematics, refer to \cite{edsjoe_neutrinos_2017}. After generation in the Sun, neutrinos are also propagated to a distance of 1 AU from the Sun, assuming standard neutrino oscillations with parameters from world best-fit values \cite{Esteban_2017} and assuming normal mass ordering. 

The spatial distribution of the neutrinos incoming to the detector are simulated following three different shapes for the Sun: {\it a)} point-like, i.e., SA$\nu$s are emitted from a single point in the sky coincident with the Sun centre; {\it b)} filled disk shape, with neutrinos produced uniformly from a disk of $0.27^\circ$ radius; {\it c)} ring-shaped, in which the outcoming neutrinos are yielded close to the Sun surface and orthogonally with respect to the direction to the Sun centre (inner radius of $0.26^\circ$ and outer radius of $0.27^\circ$).

\begin{figure}[tbh]
	\centering	
	\includegraphics[width=0.75\textwidth]{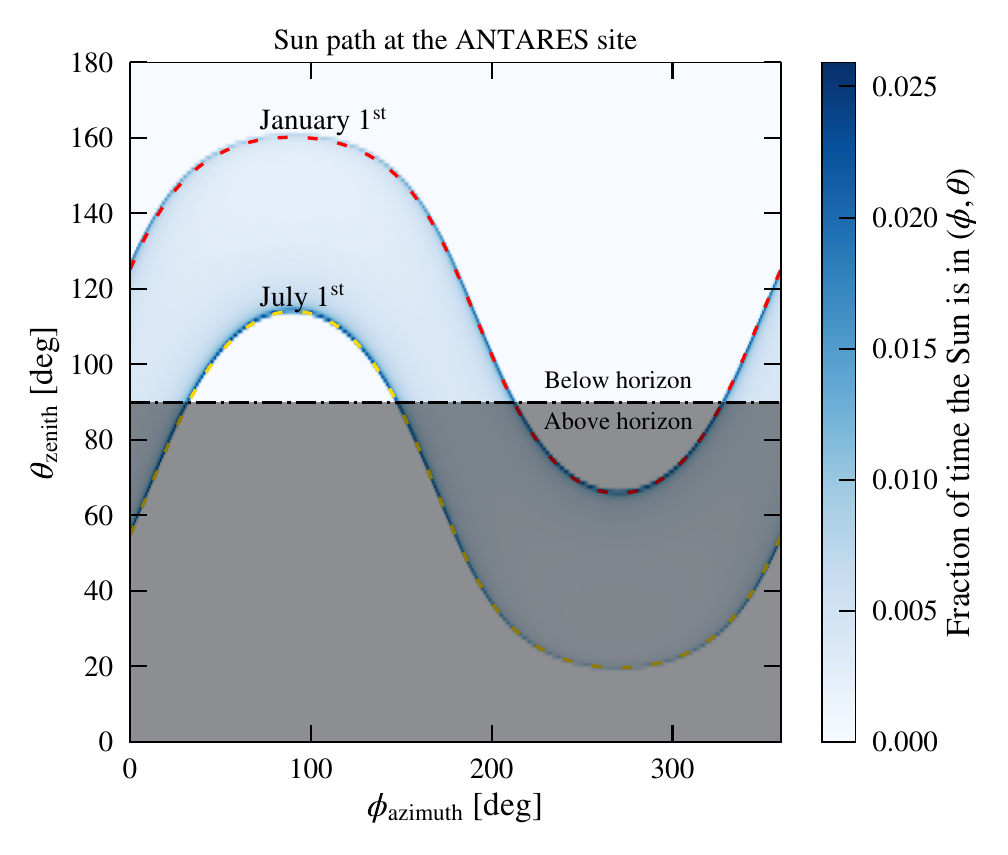}
	\caption{Solar path for the 2008--2018 period as seen from the ANTARES site. The red (yellow) dashed line shows the solar path on January (July) $1^\text{st}$, as a reference. The shaded area corresponds to the position of Sun above horizon, which represents the excluded region in this search.}
	\label{fig:Sun_path}
\end{figure}

The reference case selected for this analysis uses the \emph{H3a} CR model with the \emph{Ser+Stein} solar density profile, and considers the Sun as a \emph{point source} (this set of choices will be referred to as the {\it baseline} case). 
As we focus on long tracks in the detector, the only flavour that is considered at the detector location is the $\nu_\mu$\footnote{Here and in the following, both neutrinos and antineutrinos are consider in the symbol $\nu_\mu$.}. 
The signal in the detector consists of muons arising from the CC interactions close to the detector instrumented volume due to the considered SA$\nu_\mu$ flux models, in a period equivalent to the 11 years of ANTARES data. The simulation includes the fact that the Sun is a moving source in the local sky coordinate system, therefore the signal of SA$\nu_\mu$ is expected to be found along the solar path (figure \ref{fig:Sun_path}). When the Sun is below the horizon (i.e., $\theta_\text{zenith}>90^\circ$), events that are upward going in the detector can be produced. Due to the large flux of atmospheric muons, this study is restricted to events that are reconstructed in local coordinates as upward going.

\begin{figure}[tbh]
	\centering
	\includegraphics[width=0.7\textwidth]{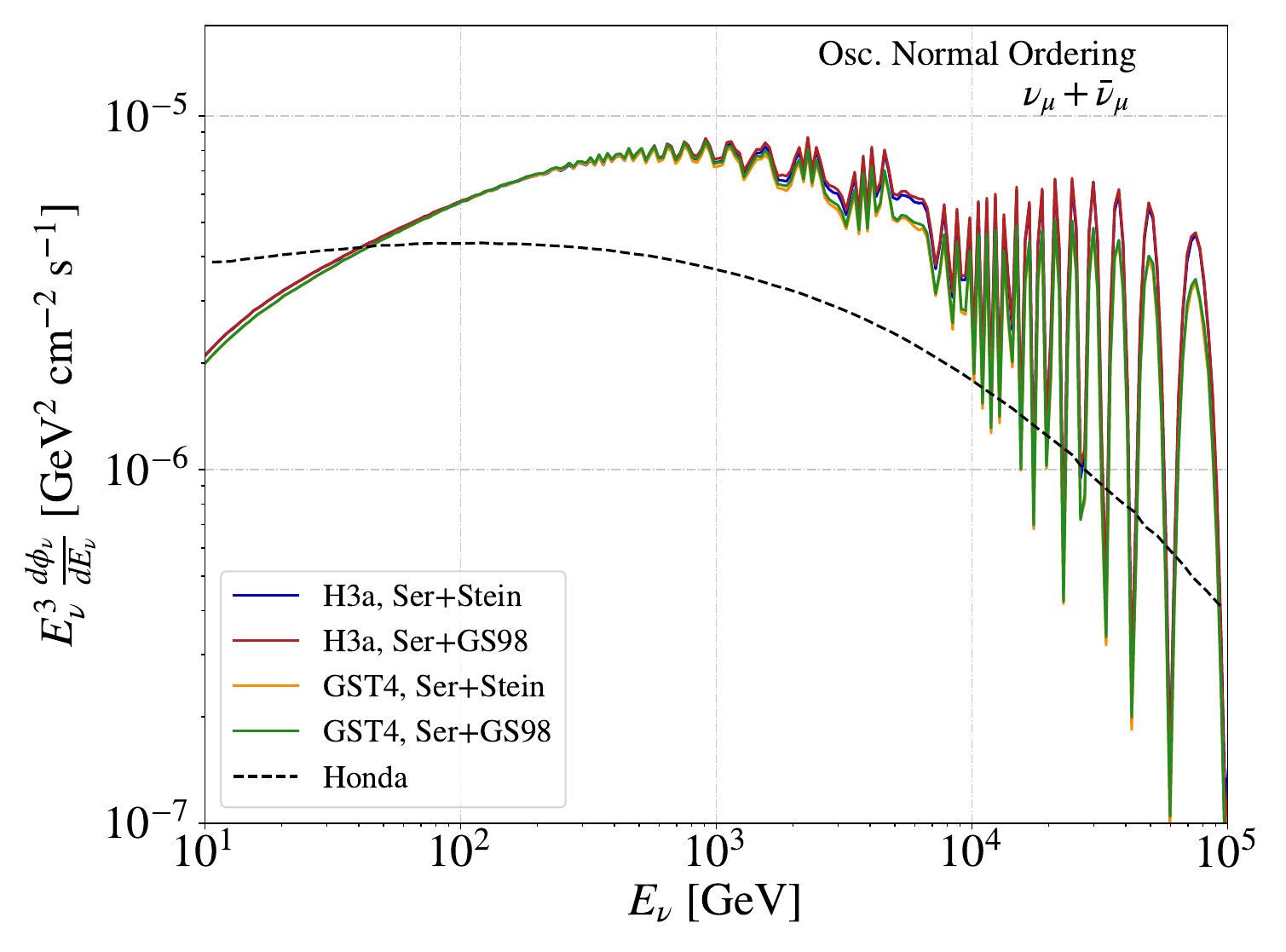}
	\caption{Muon neutrino energy fluxes, from the four different models tested in this work (see details in the text), after neutrino propagation, at the Earth position (color lines). Normal ordering and neutrino oscillation parameters from the world best-fit values are assumed. The Honda flux for Earth atmospheric muon neutrinos is shown for comparison (black dashed line) \cite{Honda_2007}. Fluxes are integrated over the solar solid angle $\Omega_\odot \simeq 6.8 \times 10^{-5}$ sr.}
	\label{fig:SAV_flux_vs_honda}
\end{figure}

Figure \ref{fig:SAV_flux_vs_honda} shows the SA$\nu_\mu$ spectra at Earth corresponding to the four tested combinations of CR flux and solar density models.
For comparison, the figure includes the Earth atmospheric $\nu_\mu$ flux according to \textit{Honda} \cite{Honda_2007}. These neutrinos are produced with the same CR flux and interaction models as the SA$\nu$, but the CR interactions occur with nuclei of the atmosphere. 
In the figure, the Earth atmospheric $\nu_\mu$ are integrated over the Sun solid angle of $\Omega_\odot \simeq 6.8 \times 10^{-5}$ sr.
As expected, since the solar atmosphere is less dense than that of the Earth, secondary mesons produced by the CR interaction in the Sun external regions are more likely to decay than to interact. Thus, 
 the Honda flux dominates for neutrino energies below $\sim 40$ GeV, while the SA$\nu$ flux prevails for $E_\nu \gtrsim 40$ GeV. 


\subsection{Background generation}
The main physics background for the present study is due to atmospheric neutrinos and atmospheric muons. 
Although they can be generated in the detector using simulation tools, as described in \cite{albert_monte_2021}, given the small expected contribution of the signal in the overall data set, the background rate
is estimated directly from the measured data.
Each reconstructed event is identified by its direction in local coordinates $(\theta_\text{zenith},\ \phi_\text{azimuth})$ and arrival time. From this information, the location on the celestial sphere, right ascension (RA) and declination $\delta$, are derived. The background rate is described as a function of the declination only. Due to the Earth's rotation and a sufficiently uniform exposure, the background is considered independent of RA.
Real events collected during 11 years of ANTARES data (2008--2018) were scrambled in RA, i.e., real measured RA angles were replaced with a random number between $0$ and $2\pi$. 


\subsection{Detector response}
After the generation of the signal (muons induced by CC SA$\nu$ interactions), particles are propagated and tracked through the detector, and Cherenkov photons are simulated and propagated to the optical modules. Finally, the data acquisition system is simulated. The environmental conditions, bioluminescence processes and sea current changes to which an undersea neutrino telescope is exposed, may affect the trigger and data acquisition system. In order to reproduce the detector response under these conditions as accurately as possible, a Monte Carlo \emph{run-by-run} strategy \cite{albert_monte_2021} is followed, in which events are simulated according to the corresponding detector state. 
After the generation of the detector hits, simulated events follow the same data acquisition chain (\textit{filtering}) of real data.

\subsection{Reconstruction and event selection\label{sec:reco_sel}}
After data filtering, the direction and energy of each triggered event (both in data and Monte Carlo) can be reconstructed from the positions and times of photomultiplier hits by different algorithms. 
In this work, the multi-line reconstruction fit employed in the search for point-like cosmic sources has been used \cite{albert_first_2017, adrianmartinez_search_2012}. 
In addition to the reconstructed direction in local coordinates, $(\theta_\text{zenith},\ \phi_\text{azimuth})$, and the number of hits passing a pre-defined threshold condition, $N_\textnormal{hit}$, the algorithm provides also two quality parameters.
The first one, denoted as $\Lambda$, is a maximum likelihood estimator 
that describes the quality of the reconstruction. The second is a parameter, referred to as $\beta$, related to the angular uncertainty on the reconstructed muon direction. Finally, the number of hits used by the reconstruction algorithm, $N_\textnormal{hit}$, is employed as a proxy for the energy of the event. Details for this algorithm are given in \cite{AdrianMartinez2013}.

Given the excellent angular resolution of ANTARES for track-like events, and the small angular size of the Sun seen from Earth ($\theta_\odot  \sim 0.5^{\circ}$), most of the sensitivity to SA$\nu$ comes from the $\nu_\mu$ CC channel. The event selection in this work follows the criteria established for the pointlike source search with the ANTARES neutrino telescope \cite{albert_first_2017, adrianmartinez_search_2012}. 
A cut on the local zenith angle, $\theta> 90^\circ$, is necessary to remove the huge background of atmospheric muons. Similarly, cuts on the parameters related to the quality of the reconstruction algorithm, $\Lambda>-5.2$ and $\beta<1^\circ$, are needed to obtain a sample of well reconstructed neutrino candidates and to reject as much as possible the background of atmospheric muons mis-reconstructed as upgoing.

With this set of cuts, the background sample is mostly due to atmospheric neutrinos: the contribution estimated from Monte Carlo is  $\nu_\mu \text{ CC} \simeq 84.7\%$, atmospheric muons $\simeq 14.8\%$ and $\nu$ NC + $\nu_e$ CC $\simeq 0.5\%$.
The total data sample after the aforementioned cuts consists of $7 071$ tracks, collected during 11 years of ANTARES data (2008--2018) with a livetime of 3022 days.




\section{Analysis\label{sec:Analysis}}

\subsection{Likelihood function\label{sec:like}}

\begin{figure}[tbh]
	\centering
\includegraphics[width=0.49\textwidth]{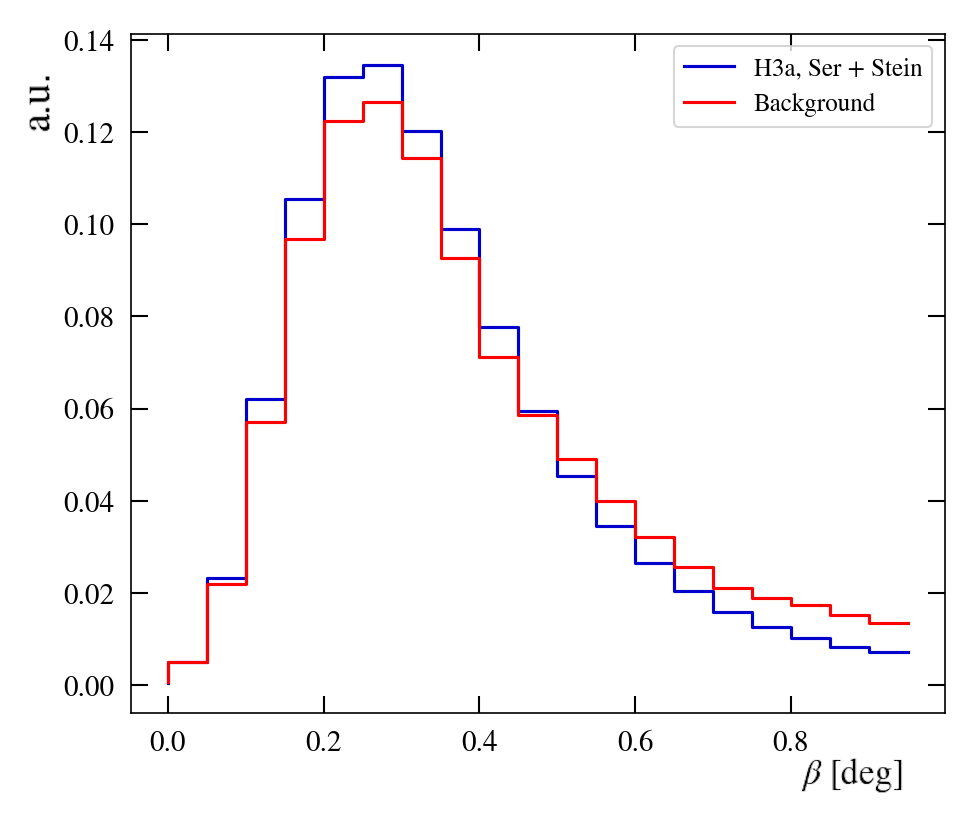}
\includegraphics[width=0.49\textwidth]{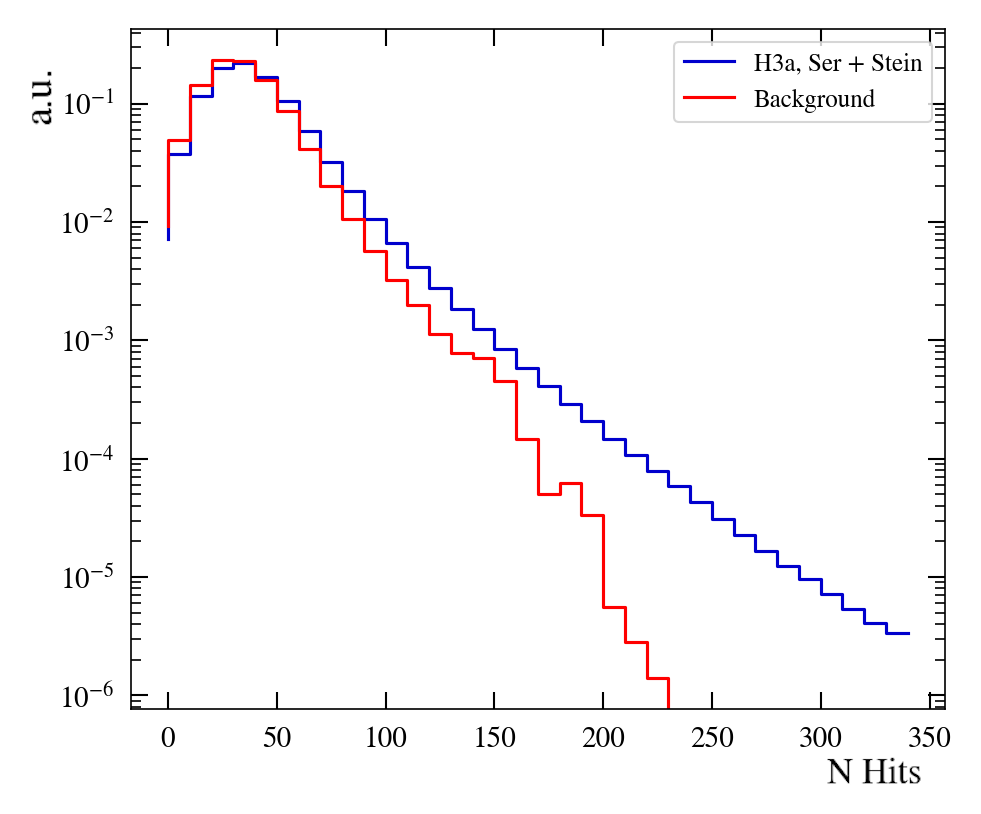}
\caption{Distribution of the $\beta$ (left) and $N_\text{hit}$ (right) parameters for the signal \textit{baseline} model (blue histogram) and for the expected background due (mainly) to Earth atmospheric neutrinos. The selection cuts defined in section \ref{sec:reco_sel} have been applied to both sample, i.e., $\Lambda > -5.2$, $\beta<1^\circ$ and $\theta> 90^\circ$.}
\label{fig:beta_nhit}
\end{figure}

The present study searches for an excess of events from SA$\nu$ with respect to the Earth atmospheric neutrinos and atmospheric muons. It relies on the determination of neutrino candidates from the Sun direction, using the number of hits as a proxy for their energy.
Discriminating variables, such as the reconstructed angular distance to the source, $\Psi_\odot$, the angular uncertainty on the reconstructed muon direction, $\beta$, and the number of hits used for the track reconstruction, $N_\textnormal{hit}$ as the energy proxy, are used to identify the signal. 
Differences on the distribution of the $\beta$ and $N_\textnormal{hit}$ variables for signal and background are visible in figure \ref{fig:beta_nhit}.
The normalised distributions of these variables are used to construct the probability density functions (PDFs) in an extended likelihood function: 
\begin{equation}
	\small
	\mathcal{L}(n_\text{sig}) = e^{-(n_\text{sig} + n_\text{bkg})} \prod_i^{N} \left[n_\text{sig} \cdot \mathcal{S}(\Psi_{\odot,i}, \beta_i, N_{\text{hit},i}) + n_\text{bkg} \cdot \mathcal{B}(\Psi_{\odot,i}, \beta_i, N_{\text{hit},i}) \right].
	\label{eq:likelihood}
\end{equation}

\begin{figure}[tbh]
	\centering
\includegraphics[width=0.49\textwidth]{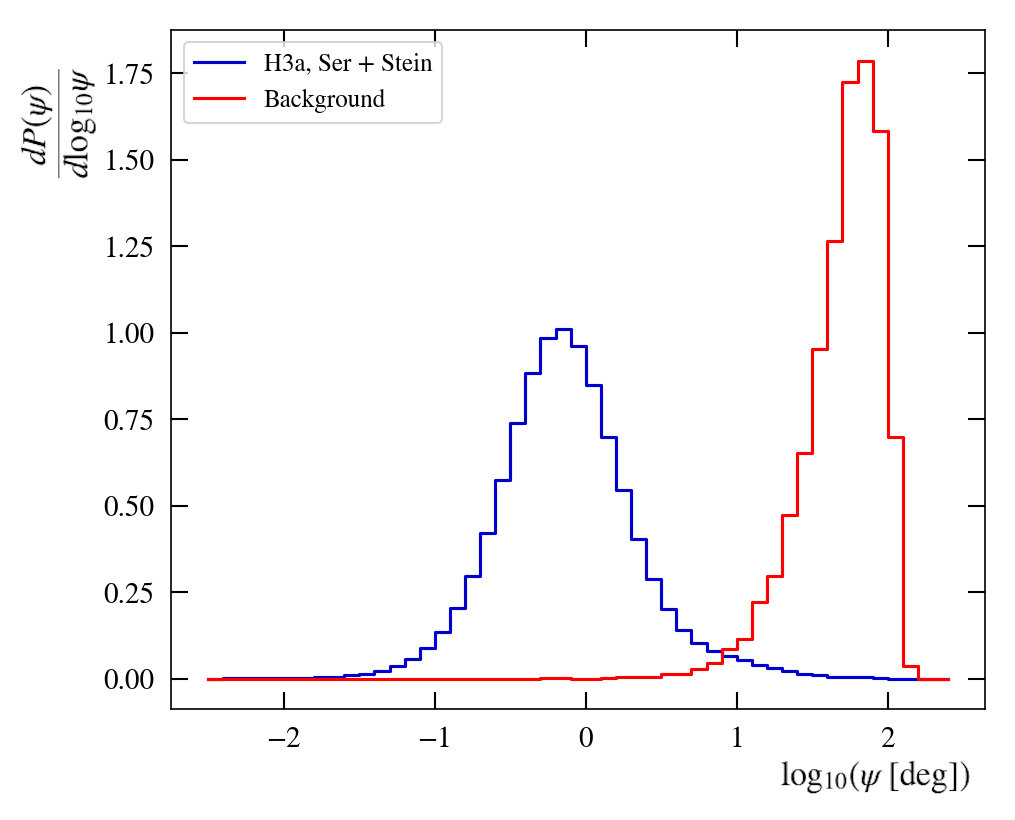}
\includegraphics[width=0.49\textwidth]{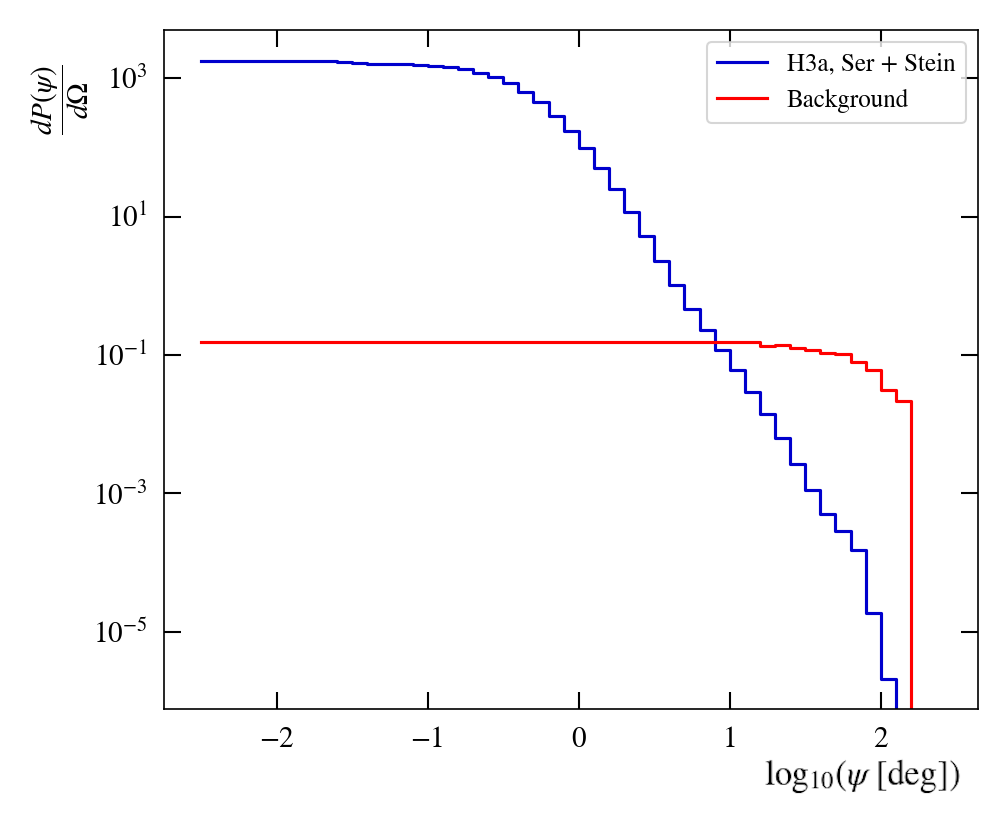}
\caption{Signal (blue) and background (red) PDFs used as inputs in the likelihood function (eq.~\ref{eq:likelihood}). The PDFs of the angular distance between the direction of the reconstructed track and the position of the center of the Sun ($\Psi_\odot$) are plotted before (left) and after (right) normalising to the covered solid angle.}
\label{fig:sig_bg_pdfs}
\end{figure}

The signal ($\mathcal{S}$) and background ($\mathcal{B}$) PDFs in eq.~\ref{eq:likelihood} are shown in figure \ref{fig:sig_bg_pdfs}. They are built from simulated events assuming the {\it baseline} SA$\nu$ flux and the background obtained with the scrambled data set, respectively. 
To evaluate the signal significance, a large set of \emph{pseudo-experiments} (PE), or \emph{skymaps}, are generated injecting a number of signal events, $n_\text{sig}$, according to the signal PDF, over the total number of detected events in the data sample $N = n_\text{sig} + n_\text{bkg}$, where the $n_\text{bkg}$ represents the expected number of background events.

The excess of events in the solar path is searched for using an unbinned likelihood method \cite{neyman_outline_1937, albert_first_2017}. The likelihood maximisation process runs over the total number of reconstructed events within a Region of Interest (RoI) of $30^\circ$ around the Sun. Due to the small extension of the source, it is possible to constrain the search to this RoI, preserving a good number of reconstructed events without missing information and speeding up the maximisation process. The RoI is chosen to be 30$^\circ$ to get a statistically significant sample of events to perform the likelihood analysis. The outcome of the maximisation is the number of signal events, $\hat{n}_\text{sig}$, which maximises the likelihood for each skymap.

The significance of an observation of a given number of signal events $\hat{n}_\text{sig}$ is evaluated through a test statistic, TS, defined as the ratio between the maximum and the background-only likelihoods:

\begin{equation}
\text{TS} = \log_{10}\left(\dfrac{\mathcal{L}(\hat{n}_\text{sig})}{\mathcal{L}(0)} \right).
\label{eq:TS}
\end{equation}

\begin{figure}[tbh]
	\centering
	\includegraphics[width=0.75\textwidth]{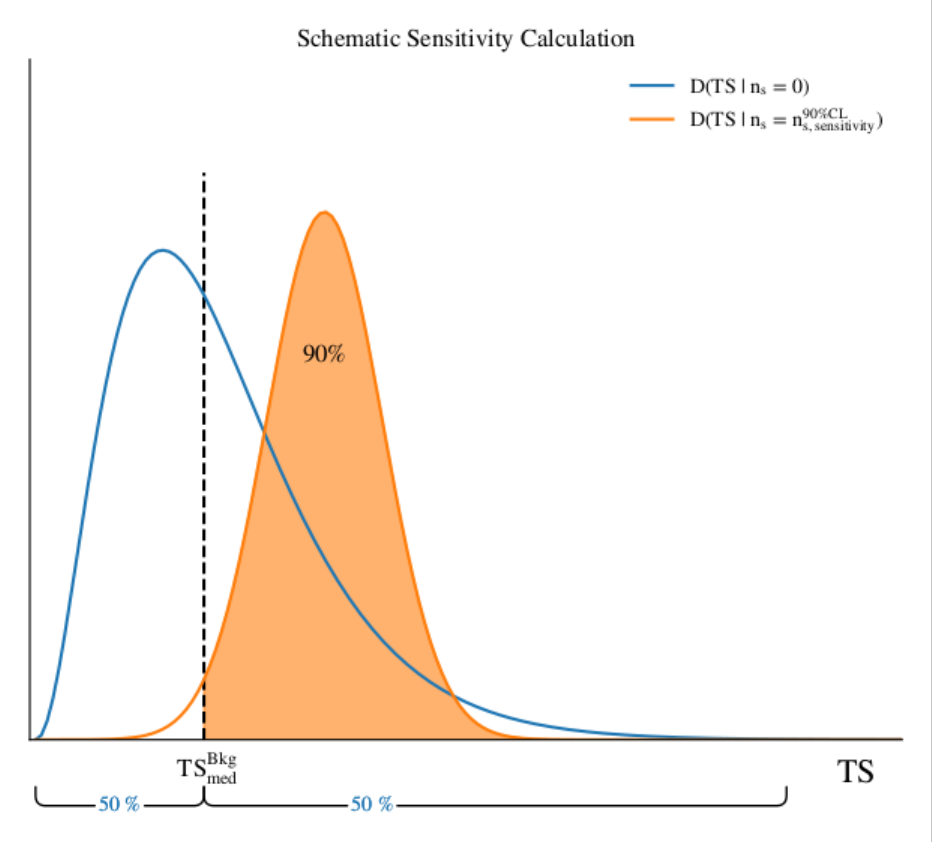}
	\caption{Schematic of the sensitivity calculation. The blue curve represents the TS distribution for zero number of injected signal (background-only hypothesis). The orange curve is the TS distribution for a number of signal events ($n_s$) injected on the skymaps (signal hypothesis), corresponding to the sensitivity. This curve has the 90\% of the distribution over the median TS of the background distribution. The upper limit is computed similarly, but instead of comparing with the TS of the median of the background distribution, the measured TS is used.}
	\label{fig:schematics}
\end{figure}

The sensitivity of the detector is defined as the 90\% CL median upper limit, which is computed by comparing the background TS distribution to the signal plus background TS distributions (see figure~\ref{fig:schematics}) \cite{neyman_outline_1937}. The computation of the sensitivity is done following the next equation

\begin{equation}
	n_\text{sig} = n_{\text{sig}, \text{ sensitivity}}^\text{90\%CL} 
	\hspace{0.25cm} \text{if} \hspace{0.25cm}
	\displaystyle\int^{\infty}_{\text{TS}^\text{Bkg}_{\text{med}}} D(\text{TS} | n_\text{sig}) \text{dTS} = 90\%
\label{eq:missing}
\end{equation}

The $D(\text{TS} | n_\text{sig})$ term in eq.~\ref{eq:missing} represents the TS distribution for a given number of signal events, $n_\text{sig}$, over which the integration is done. The TS$_\text{med}^\text{Bkg}$ is the median of the background TS distribution. The sensitivity is set to $n_\text{sig} = n_{\text{sig}, \text{ sensitivity}}^\text{90\%CL}$ for the TS distribution that fulfills the condition in eq.~\ref{eq:missing}.

Following the same approach, the measured upper limit is computed from the observed TS after unblinding the data.  If the observed TS is below the median of the TS background distribution, the upper limit is set equal to the sensitivity. The \emph{p-value} for the upper limit is computed comparing the measured TS with the background-only TS distribution as

\begin{equation}
	\text{\emph{p-value}} = \displaystyle\int_{\text{TS}_\text{measured}}^\infty D(\text{TS}|n_\text{sig} = 0) \text{dTS}
\end{equation}

If no signal is observed, a limit on the neutrino flux is computed from the limit on the number of signal events, $n_\text{sig}^{90\% \text{CL}}$, according to the following expression:

\begin{equation}
\dfrac{d \Phi^\text{90\%\ CL}_{\nu_\mu} (E)}{dE} =
\dfrac{n^\text{90\%\ CL}_\text{sig}}{\bar{n}^\text{theor}_\text{sig}}
\dfrac{d \Phi^\text{theor}_{\nu_\mu} (E)}{dE}=
C_{90} \cdot
\dfrac{d \Phi^\text{theor}_{\nu_\mu} (E)}{dE} 
.
\label{eq:flux_sensi}
\end{equation}

\begin{figure}[tbh]
	\centering
	\includegraphics[width=0.65\textwidth]{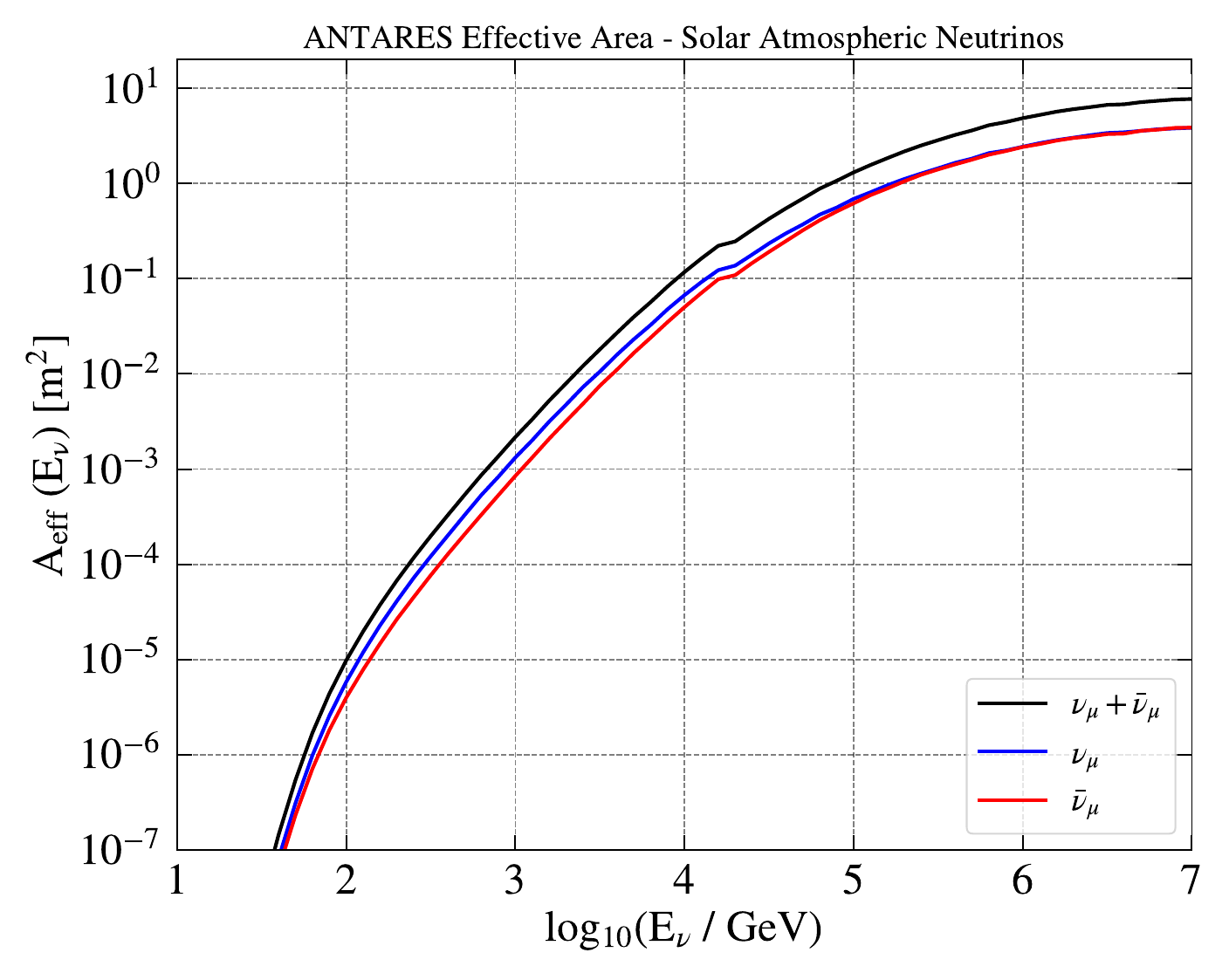}
	\caption{ANTARES effective area for $\nu_\mu$, $\bar{\nu}_\mu$ and the sum of the two, arriving from the Sun direction, 
after the selection cuts ($\Lambda>-5.2$, $\beta<1^{\circ}$ and $\theta > 90^{\circ}$). The effective area depends on the local zenith angle $\theta$ and the plot is made by averaging the Sun positions during the considered period (2008--2018).}
	\label{fig:Aeff}
\end{figure}

The first term in equation \ref{eq:flux_sensi} corresponds to the flux upper limit. The second and third terms represent the theoretical flux model multiplied by a scale factor, $C_{90}$,  defined as the ratio between $n_\text{sig}^{90\% \text{CL}}$ and the expected number of signal events $\bar{n}^\text{theor}_\text{sig}$ detected. The expected number of signal events $\bar{n}^\text{theor}_\text{sig}$ is computed in the following way:

\begin{equation}
\bar{n}^\text{theor}_\text{sig} =
T \displaystyle\int{
\sum_{l\in \nu_\mu, \bar{\nu}_\mu}{
\left( \dfrac{d \Phi^\text{theor}_{l} (E)}{dE} A^\text{eff}_{l} (E) \right) dE}},
\label{eq:nsig_expected}
\end{equation}

\noindent where $T$ is the livetime of data taking, $A^\text{eff}$ is the ANTARES effective area for this analysis (see figure \ref{fig:Aeff}), and ${d \Phi^\text{theor}_{l} (E)}/{dE}$ is the theoretical flux model.

In the considered baseline scenario of SA$\nu$s, the expected number of signal events for the 3022 days of livetime is $\bar{n}_\text{sig}^\text{theor} = 0.37$.

\subsection{Systematic uncertainties}
The systematic uncertainties arise from the signal simulation and from the detector response. The former includes the differences on the $\nu_\mu$ spectra from different CR models and solar density profiles (as shown in figure \ref{fig:SAV_flux_vs_honda}) up to a 30\% uncertainty in the neutrino production flux as discussed in \cite{edsjoe_neutrinos_2017, Fedynitch_2012}, and those arising from different spatial distribution when the three different shapes for the Sun (point-like, filled disk and ring-shaped) are assumed. The latter includes effects on the detector absolute pointing accuracy, the angular resolution for upgoing muon tracks, and the detector acceptance.

The effect on these systematic uncertainties are considered with respect to the selected reference case mentioned in section \ref{sec:data_set}, i.e., the baseline scenario obtained with the \emph{H3a} CR model, the \emph{Ser+Stein} solar density profile, the Sun as a point-like source, and nominal detector performances.
For the signal, normal mass ordering and neutrino oscillation parameters from best-fits values are assumed \cite{edsjoe_neutrinos_2017, Esteban_2017}. The analysis is optimised for $\nu_\mu$ CC interactions yielding muons crossing the detector. 
The inclusion of systematics uncertainties will worsen the sensitivity as determined for the baseline scenario case. 

Systematics due to the detector response includes different effects, as studied in previous ANTARES papers \cite{albert_first_2017, adrianmartinez_search_2012, Albert_2018}.
The ANTARES absolute pointing accuracy uncertainty was determined using the Moon \cite{ANTARES:2018gll} and Sun \cite{collaboration_observation_2020} shadows. 
To consider the possible displacement of the Sun from the nominally expected reconstructed position, randomly generated offsets have been added to the $\phi_\text{azimuth}$ and $\theta_\text{zenith}$ variables describing the centre of the Sun (Figure \ref{fig:Sun_path}) of simulated events. The offsets are generated according to two Gaussian distributions with the width according to \cite{collaboration_observation_2020}.
The angular resolution of the track reconstruction algorithm can be affected by the accuracy of the detected hit times.
A smearing of these times was performed in simulations, leading to up to 15\% degradation on the angular resolution for reconstructed muons. 
Uncertainties on the knowledge of water properties (attenuation, scattering length) and optical modules efficiencies induces an uncertainty on the detector effective area (corresponding to the detector acceptance in neutrino telescopes). 
To constrain this systematic uncertainty, a comparison is performed between the events obtained with nominal detector parameters and simulations in which water properties, or the efficiency of the optical modules, are varied according to the known uncertainties on the values used for the simulation. The corresponding effect induces variations up to 15\% on the detector acceptance in the considered energy range.

When considering all these uncertainties, it is found that the median sensitivity at 90\% CL would worsen by about 5\% with respect to the baseline scenario. 
An effect of less than 2\% arises when considering the other combinations of CR model and solar density profile.
Finally, it turns out that the largest systematic effect on sensitivity's uncertainty arises from the different shapes of the Sun. The different values of sensitivities, $n^\text{90\% CL}_\text{sig, sens}$, represented by the number of events obtained by eq. \ref{eq:missing} for the three Sun shapes considered, are included in the first column of Table \ref{tab:sensi}.




\section{Results and conclusions\label{sec:Results} }

As described in section \ref{sec:like}, the excess of events in the solar path is searched for through an unbinned likelihood function that uses events contained in a RoI of $30^\circ$ around the Sun centre. The percentage of expected signal falling inside the RoI is $\sim 99.6\%$.
The expected number of background events is 470.
When opening the (2008--2018) real data set, the number of detected events in the data sample contained in the RoI is 461. 
The unbinned likelihood function yields no excess of SA$\nu$ signal over the expected background in the 11 years of analysed data.

\begin{table}[h]
    \centering
    \begin{tabular}{lcc||cc}
Sun Shape & $n^\text{90\% CL}_\text{sig, sens}$ & Ratio & $n^\text{90\% CL}_\text{sig, up-lim}$  & \textit{p-val} \\ 
        \midrule\midrule
Point-like & 2.70 & 1.00& 3.15 & 0.41 \\
Filled disk & 2.80 & 1.04 & 3.25 & 0.43 \\
Ring-shaped & 3.45 & 1.28 & 3.45 & 0.50 \\
\end{tabular}
\caption{First column: Sensitivities in terms of $n^\text{90\% CL}_\text{sig, sens}$ for the three different Sun shapes considered in the SA$\nu$ \textit{baseline} model. The second column is the ratio of the sensitivities with respect to the point-like case.
The third and fourth columns report the corresponding 90\% CL upper limits and the \textit{p-values} corresponding to the quoted upper limits for the real data set.}
    \label{tab:sensi}
\end{table}

\begin{figure}[tbh]
	\centering
	\includegraphics[width=0.7\textwidth]{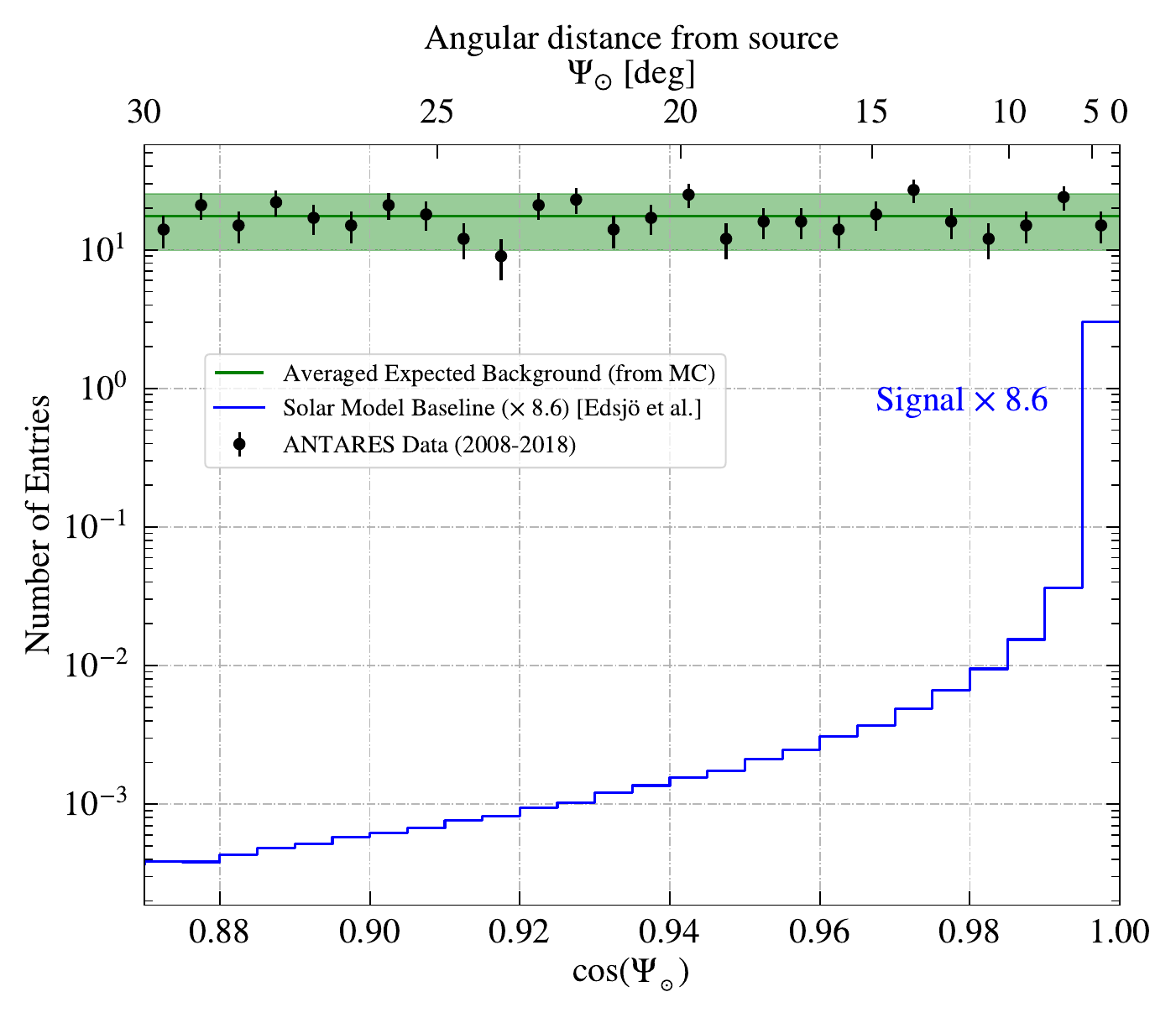}
	\caption{Number of detected events (black dots) as a function of the reconstructed angular separation $\Psi_\odot$ with respect to the Sun centre. The expected signal, in blue, is scaled up by a factor 8.6. The expected background (green line) is shown with a $2\sigma$ band along the data (black dots).}
	\label{fig:cos_psi}
\end{figure}

In the $baseline$ scenario, the 90\% CL upper limit obtained after analysing the unblinded data is $n_\text{sig, up-lim}^\text{90\% CL} = 3.15$, corresponding to a flux scale factor of $C_\text{90}= 8.6$. 
This value of the scale factor indicates the possibility of excluding at 90\% CL the tested model. A value smaller than one will directly constrain the model. In this study the flux that can be constrained would be 8.6 times larger than the one of the baseline model. 
Figure \ref{fig:cos_psi} shows the distribution of the events within the RoI of $30^\circ$ around the Sun, for the expected signal (blue histogram) and background (green line), alongside the observed data (black dots). The intensity of the signal is magnified by a factor 8.6 for comparison. 
In table \ref{tab:sensi} the upper limits on $n^\text{90\% CL}_\text{sig, up-lim}$  obtained after data unblinding for the three Sun shapes considered, as well as the corresponding \emph{p-values}, are reported. The first column contains the evaluated sensitivities.
%

\begin{figure}[tbh]
	\centering
	\includegraphics[width=0.8\textwidth]{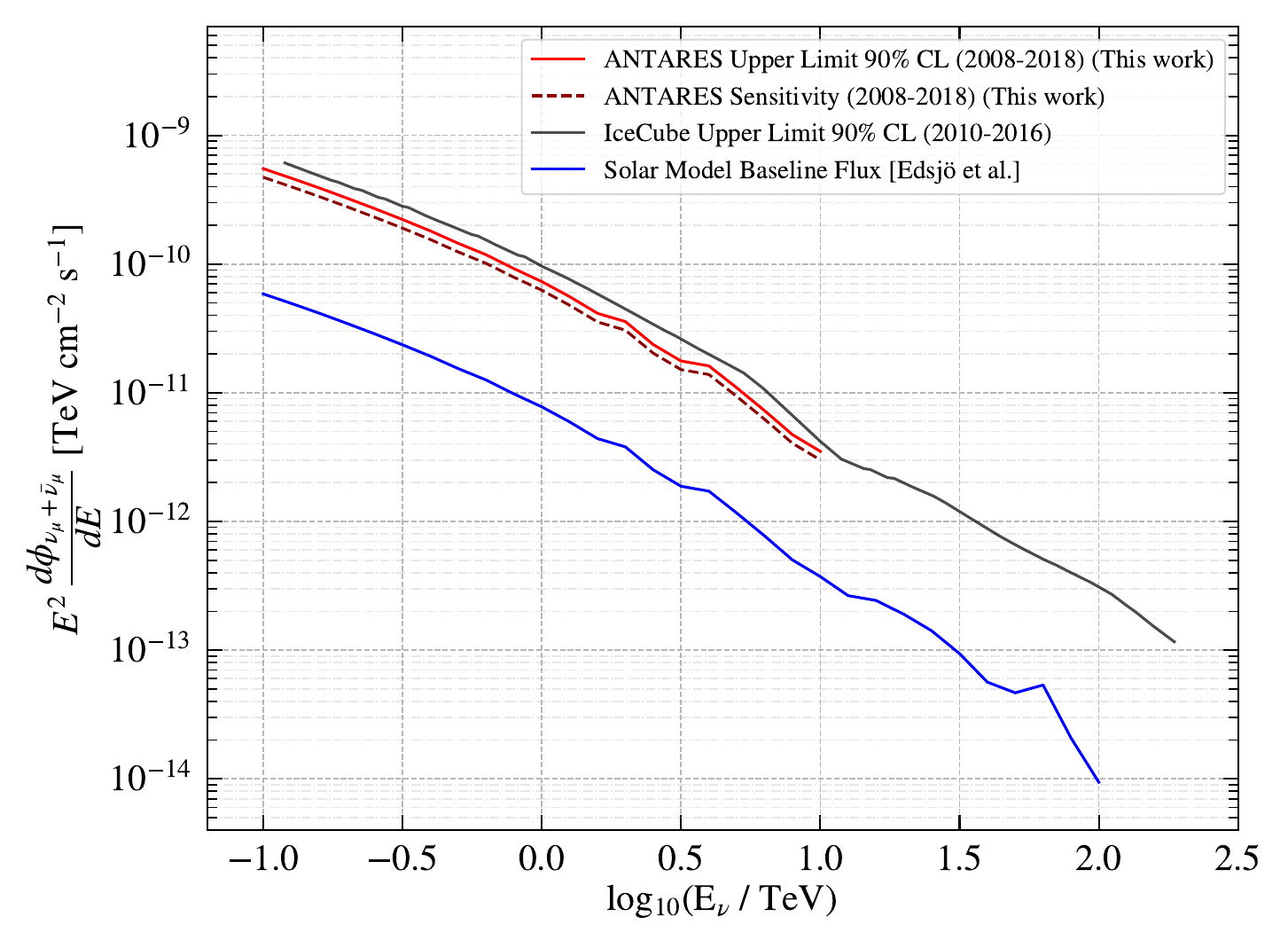}
	\caption{ANTARES upper limit (solid red) and sensitivity (dashed red) for 11 years of data, assuming the Sun as a point-like source for the baseline model \textit{H3a-Ser+Stein} (solid blue line). For comparison, the current 6 years IceCube upper limit \cite{IC_solarnu_2021} is also shown (solid black line). The ANTARES limit and sensitivity lines expand in the energy range which contains 90\% of the expected number of events.}	
	\label{fig:ant_upp_lim}
\end{figure}

Figure \ref{fig:ant_upp_lim} presents the 90\% CL upper limit (solid red line) on the $baseline$ SA$\nu$ flux as a function of the neutrino energy obtained by the ANTARES detector using 3022 days of livetime. The corresponding sensitivity is also indicated as a dotted red line. The limit covers the energy range which contains 90\% of the expected number of SA$\nu$ events. The theoretical flux model (solid blue line) and the upper limits obtained by the IceCube collaboration (solid black line) \cite{IC_solarnu_2021} are included in the figure for comparison.
The \emph{GST4} cosmic ray model and the \emph{Ser+GS98} solar density profile have been tested in combination with the models used in the baseline scenario (see figure \ref{fig:SAV_flux_vs_honda}), and the results are within a 2\% difference with respect to the values shown in table \ref{tab:sensi}.


After analysing 11 years of ANTARES data, corresponding to 3022 days of total livetime, with an unbinned likelihood method, using three different sun shapes, no signal evidence of SA$\nu$s is observed. As a result, 
a 90\% CL upper limit on the flux of 7$\times$$10^{-11}$ TeV$^{-1}$cm$^{-2}$s$^{-1}$ at $1$ TeV is established. 
In this context, the advent of the new neutrino telescope in the Mediterranean Sea (the KM3NeT detector \cite{adrianmartinez_letter_2016}) that will improve both the mentioned requirements would represent a significant progress toward the important observation of SA$\nu$s.

\section*{Acknowledgements}
The authors acknowledge the financial support of the funding agencies:
Centre National de la Recherche Scientifique (CNRS), Commissariat \`a
l'\'ener\-gie atomique et aux \'energies alternatives (CEA),
Commission Europ\'eenne (FEDER fund and Marie Curie Program),
Institut Universitaire de France (IUF), LabEx UnivEarthS (ANR-10-LABX-0023 and ANR-18-IDEX-0001),
R\'egion \^Ile-de-France (DIM-ACAV), R\'egion
Alsace (contrat CPER), R\'egion Provence-Alpes-C\^ote d'Azur,
D\'e\-par\-tement du Var and Ville de La
Seyne-sur-Mer, France;
Bundesministerium f\"ur Bildung und Forschung
(BMBF), Germany; 
Istituto Nazionale di Fisica Nucleare (INFN), Italy;
Nederlandse organisatie voor Wetenschappelijk Onderzoek (NWO), the Netherlands;
Executive Unit for Financing Higher Education, Research, Development and Innovation (UEFISCDI), Romania;
Ministerio de Ciencia, Innovaci\'{o}n, Investigaci\'{o}n y
Universidades (MCIU): Programa Estatal de Generaci\'{o}n de
Conocimiento (refs. PGC2018-096663-B-C41, -A-C42, -B-C43, -B-C44)
(MCIU/FEDER), Generalitat Valenciana: Prometeo (PROMETEO/2020/019),
Grisol\'{i}a (refs. GRISOLIA /2018 /119, /2021 /192) and GenT
(refs. CIDEGENT/2018/034, /2019/043, /2020/049, /2021/023) programs, Programa Operativo FEDER 2014-2020/Junta de Andalucía-Consejería de Economía y Conocimiento/ Proyecto A-FQM-053-UGR18, La Caixa Foundation (ref. LCF /BQ /IN17 /11620019), EU: MSC program (ref. 101025085), Spain;
Ministry of Higher Education, Scientific Research and Innovation, Morocco, and the Arab Fund for Economic and Social Development, Kuwait.
We also acknowledge the technical support of Ifremer, AIM and Foselev Marine
for the sea operation and the CC-IN2P3 for the computing facilities.
\bibliographystyle{JHEP}
\bibliography{AllBib}

\end{document}